# High harmonic interferometer:
# For probing sub-laser-cycle electron dynamics in solids

K. Uchida[1,*] and K. Tanaka[1,]

[1] *Department of Physics, Graduate School of Science, Kyoto University, Kyoto, Kyoto 606-8502, Japan*

**Abstract**

High harmonic emissions from crystalline solids contain rich information on the dynamics of electrons driven by intense infrared laser fields and have been intensively studied owing to their potential use as a probe of microscopic electronic structures. Especially, the ability to measure the temporal response of high harmonics may allow us to investigate electron dynamics directly in quantum materials. However, this most essential aspect of high harmonic emissions has been challenging to measure.

Here, we propose a simple solution for this problem: a high harmonic interferometer, where high harmonics are generated in each of the path of a Mach-Zehnder interferometer and an interferogram of them is captured. The high harmonic interferometer allows us to achieve a relative time resolution between the target and reference high harmonics of less than 150 attoseconds, which is fine enough to track sub-cycle dynamics of electrons in solids. By using high harmonic interferometrer, we succeeded in capturing the real time dynamics of Floquet states in $WSe_2$, whose indirect signature had so far been caught only by time-averaged measurement. Our simple technique will open a door to attosecond electron dynamics in solids.

## I. INTRODUCTION

Driving of electrons in condensed matter with an intense infrared (IR) laser field provides us with a fascinating playground in which to study intriguing phenomena far from equilibrium [1]. When IR driving is strong enough to induce electron motion across a full Brillouin zone in crystalline solids, the electrons exhibit sub-laser-cycle electron dynamics characterized by the typical energy scale of the system, e.g., the hopping energy needed to jump to a neighbor site, the bandgap energy, or the Coulomb interaction. Since these energy scales are typically on the order of eV, the corresponding dynamics are expected to be on the sub-femtosecond scale. The ability to observe such sub-laser-cycle electron dynamics in solids under an IR field allows us to access to many-body electronic structures that cannot be measured by linear optical spectroscopy in equilibrium [2].

High harmonic generation (HHG), i.e., the emission of radiation at integer multiples of the incident laser photon energy, indeed arises from sub-laser-cycle electron dynamics in the whole Brillouin zone. HHG in crystalline solids is considered to be a promising way to probe microscopic electronic properties such as band dispersion, transition dipole moment, Berry curvature, valence electron densities, atomic bonding, and so on [3-13]. In previous experimental studies, researchers mainly measured the time-averaged high harmonic (HH) intensity spectra, though it lacks HH-phase information.

To fully understand sub-cycle electron dynamics in solids, it is essential to capture the real time structure of high harmonics or its equivalent information: the HH amplitude and phase. Many numerical simulations and theoretical calculations have shown that the HH emission in the time domain directly contains information about the electron dynamics that would hard to obtain from time-averaged measurements [14-22]: the temporal structure of the HH emission can help us to understand the dominant microscopic HHG mechanisms, such as electron-hole recollisions, Kerr-type nonlinearity, Brunel harmonics, and dynamical Bloch oscillation, which have been often debated [14]. Time-domain HHG measurements are also expected to be able to capture the behavior of topologically non-trivial systems [17] or the effective dephasing due to strong correlation effects at a finite temperature [18].

Despite its importance, only a few experimental studies have attempted to measure high harmonic generation in solids in the time domain because of technical difficulties [23-25]. Pioneering studies by Hohenleutner *et al.* [24] and Langer *et al.* [25] succeeded in capturing the temporal structure of HH in a GaSe crystal in the near infrared range. They used sum frequency generation (SFG) between HH and gating pulses to reconstruct HH in the time domain. Although this scheme has helped to elucidate sub-laser-cycle electron dynamics, there are several difficulties and limitations. One is that an ultrashort gating pulse much shorter than the cycle of the driving field is needed. The pulse width of a gating pulse and a nonlinear crystal for SFG limits the time resolution of the system to about 1 femtosecond. Another problem is that a sufficiently high photon flux of HH is required for SFG detection. Here, a thick crystal is used to obtain high enough photon flux of HH emissions. However, nonlinear propagation effects in thick materials often change the fundamental properties of the high harmonics [26,27]. To observe sub-laser-cycle electron dynamics in a wide range of materials, an alternative convenient method is required.

Here, we propose a new type of high harmonic emission measurement system; the high harmonic interferometer (HH interferometer). This instrument allows us to access the charge dynamics in solids with a relative time resolution of less than 150 attoseconds. By using HH interferometry, we accessed the real-time Floquet state dynamics in 2H-WSe$_2$, where only the time-averaged signature had been observed in the previous study [28].

## II.   PRINCIPLE OF HIGH HARMONIC INTERFEROMETRY

In this section, we describe the principle of high harmonic interferometry. This instrument is designed to measure the "relative" timing of the high harmonic emission with a fine time resolution by using an interferometric technique.

In a conventional interferometer such as the Mach-Zehnder interferometer depicted in Fig. 1A, the incident light is split into two paths and combined after adding an optical delay to one of the paths. Depending on the delay, the intensity of the output light is modulated by the interference between the two paths and is called an interferogram. The Fourier transform of the interferogram represents the amplitude and phase of the light. By inserting the target sample into one of the paths, we can determine its complex refractive index from interferograms made with and without the target sample. This is the principle of Fourier transform infrared spectroscopy (FTIR).

The essential point of our idea is to generate high harmonic emissions inside the Mach-Zehnder interferometer and measure their interferograms, as shown in Fig. 1B. Incident mid-infrared (MIR) pulses are split into two paths. Then, the MIR beams are focused onto the reference and target samples, respectively, to induce high harmonics. The high harmonic emissions generated in each path are collimated and combined by adding an optical delay in the reference path. Since high harmonic generation is a coherent optical process and its amplitude and phase are determined by the temporal profile of the MIR driving field, high harmonics of the reference and those of the target are expected to interfere, as in the conventional interferometer. Denoting the time profiles of the electric fields of the high harmonic emissions from the reference and the target by $E_{\text{ref}}(t)$ and $E_{\text{tgt}}(t)$, the combined HHG intensity $I_{\text{HH}}$ is described as follows:

$$I_{\text{HH}}(\tau) = \frac{1}{2}\varepsilon_0 c \int dt \left|E_{ref}(t-\tau) + E_{tgt}(t)\right|^2$$
$$= I_{ref} + I_{tgt} + \varepsilon_0 c \int dt\, E_{tgt}(t) E_{ref}(t-\tau). \quad (1)$$

Here, $\tau$ is the optical delay between reference and target paths, $\varepsilon_0$ is the vacuum permittivity, and $c$ is the speed of light. The third term describes the interference of the high harmonic signals, and we call it an HH interferogram (Fig. 1C).

The HH-interferogram measurement is a kind of optical heterodyne detection of high harmonic emissions [29-32]. If $E_{ref}(t)$ is known, we can fully reconstruct $E_{tgt}(t)$ from the interferogram. Even if it is not known, the changes in the interferogram indicate the timing shift of the high harmonic emissions with respect to the mid-infrared driving field.

The first advantage of the HH interferometer is that we do not need any complicated optical system or an additional gating pulse; we only need the mid-infrared driving field itself. In principle, the time resolution of the system is simply determined by the stability of the interferometer. Here, the interferometric technique is widely used in modern science and industry, and a method of stabilization at an attosecond time scale has been established. The second advantage is that the HH interferometer has enough sensitivity to measure weak HHG signals thanks to the feature of the optical heterodyne detection scheme, that are difficult to detect using the SFG method in Refs. 24 and 25. As indicated in Eq. (1), the HH interferogram is proportional to both $E_{tgt}$ and $E_{ref}$. Therefore, the ratio of the interferogram's strength to the target HH intensity $I_{tgt} \propto |E_{tgt}|^2$ is given by $E_{ref}/E_{tgt}$. The signal-to-noise ratio of HH interferometry is improved by using a high-photon flux HHG source as a reference ($E_{ref} \gg E_{tgt}$). The third advantage is that we do not necessarily need the carrier envelope phase (CEP) stability of the MIR field to detect the HH interferogram. In the SFG method used in Refs. 24 and 25, the CEP stability of the driving field is crucial to observe the HH emission in the time domain. On the other hand, since the HH-interferometer allows us to measure the relative timing of the HH emission, the effect of the CEP drift on the inteferogram can be canceled out if the MIR pulse is a multi-cycle one.

In addition, a Mach-Zehnder type of interferometer allows us vary the excitation and detection conditions of the reference and target samples independently, which is difficult to do for single beam path interferometry used to measure the phase of HH emissions from gas [29-31]. This capability is helpful for measuring the target HH phase as a function of MIR intensity and polarization.

### III.  SETUP OF HH INTERFEROMETER

To demonstrate the above principle, we constructed an HH interferometer and measured the relative timing of high harmonic emissions from crystalline solids. We used a CEP-locked MIR pulse for excitation, with a center wavelength of 4900 nm and a pulse duration of 100 femtoseconds (full width of half maximum). The details of the experimental setup are described in Appendix A. Here, we chose 100-µm-thick GaSe crystal as the reference sample (top panel of Fig. 2A). As shown in a previous study, GaSe several hundreds of µm thick can generate a high photon flux HH of around 100 nJ that shows an almost instantaneous response to a MIR electric field [24]. Such a high photon flux enhances the detection sensitivity of the target HHG, and the instantaneous response of the reference HH emission allows us to determine the HH phase of the target sample. We prepared two different samples: GaSe and bulk $WSe_2$ as targets. When the target sample is the same material as the reference sample, the HH interferogram should show an instantaneous response, which allows us to evaluate the performance of the HH interferometer.

We detected the interferometric HH signals by using a spectrometer equipped with a charge-coupled device camera because typical HH emissions have a broadband spectrum with a discrete peak at the harmonic energy. In this spectrally resolved detection scheme, the Fourier transform of the interferogram is performed prior to detection, unlike in conventional interferometry such as FTIR. There are two advantages to using a spectrometer. One is that we can detect each harmonic signal separately. In the high harmonic emission, the intensity of the output signal typically decreases as the harmonic order increases. Thus, the intensity fluctuations in the lower order harmonics disrupt the detection of the higher order harmonics with a high signal-to-noise ratio if we perform the HH-interferometry without a spectrometer. Spectrally resolved measurements can remove such noise from a high harmonic detection. The second advantage is that we can save time in measuring the HH interferogram. Without a spectrometer, to obtain spectral information, it is necessary to scan the optical delay that is longer than the typical coherence time of light (e.g., the pulse duration of the mid-infrared driving field) with a sufficiently short time step. Such a long-duration scan often wastes an enormous amount of time, leading to drift or instability in the HH interferometer. In the spectrally-resolved measurement, the observed signal $I_{\mathrm{HH}}$ is transformed into

$$I_{\mathrm{HH}}(\hbar\omega,\tau) = I_{\mathrm{ref}}(\omega) + I_{\mathrm{tgt}}(\omega) + c\varepsilon_0 \left|\tilde{E}_{ref}(\omega)\tilde{E}_{tgt}(\omega)\right|\cos(\omega\tau + \Delta\phi),$$

(2)

where $\Delta\phi = \phi_{tgt}(\omega) - \phi_{ref}(\omega)$ is the phase difference between the reference and the target HHG emission. To extract the phase information $\Delta\phi$, only a single cycle of the high harmonic signal oscillation is needed. Note that there is a drawback to this detection scheme; the time window of the interferogram is limited by the inverse of the spectral resolution of the system. In our experimental setup, the spectral resolution of ~ 15 meV at 2 eV gives a sufficient time window of ~ 300 femtoseconds to examine the nonlinear optical processes occurring during the mid-infrared driving field (<100 femtoseconds).

## IV.   RESULTS
### A. HH interferometry with GaSe-GaSe combinations

Here, we perform HH interferometry by using the same material (GaSe) as a target and reference samples to evaluate the specification of the HH interferometer. Figure 2A shows HH spectra measured from 100-μm-thick GaSe (reference, orange dashed line) and 2-μm-thick mechanically exfoliated GaSe on fused silica substrate (target, blue solid line) with a 5 seconds exposure. The intensity of the even-order harmonics relative to the odd-order harmonics is much stronger for the thick reference sample, and we could not detect the even-order harmonics in the thin target sample. The enhancement of the even harmonics in the thick sample originates from propagation and cascade processes reported in the previous study [26], and it improves the detection sensitivity of the weak even-order harmonics from the target sample. Figure 2B shows the HH interferogram of the target GaSe sample as a function of the optical delay and emission energy. Here, we observe clear interference signals depending on the optical delay in the fifth and seventh harmonics. In addition, the interference signal of the sixth harmonic could be detected with an exposure time of 1 second (blue dotted line in Fig. 2C) although much weaker than the fifth and seventh harmonics. This is in contrast to the intensity measurement in Fig. 2A, where only the fifth and seventh harmonic signals are detectable. The optical homodyne technique used in HH-interferometry allows to detect such a weak signal.

Figure 2C shows the sliced interference profiles of the fifth, sixth and seventh harmonic emissions. We confirmed that the oscillation frequencies correspond to the emission frequencies of the high harmonics. In the fifth and seventh harmonics, the visibility of the interference is approximately 70 % of the ideal value estimated from the reference and target HHG spectra. The spectral resolution of the detection system and distortion in the HH-beam wavefront between the reference and target can be attributed to this discrepancy from the ideal value. The temporal drift during measurements (20 minutes) is estimated to be less than 150 attoseconds. This value is sufficiently short to investigate the sub-laser-cycle response of the electrons in the target sample. We also confirmed that the long-time stability of the HH system during repeated measurements over 6 hours is less than 700 attoseconds without any feedback (see Appendix B for a detailed description).

We can extract the phase spectrum $\Delta\phi(\omega)$ by performing a fitting of the inteferogram using Eq. (2). Figure 3 shows the extracted HH-phase and amplitude spectra of the target GaSe. The phase decreases with increasing emission photon energy. The slope of the phase is proportional to the timing of the HH emission with respect to the delay defined in Fig. 2(C), indicating that the peak of the emission in the time domain is at a later time delay. This phase profile can be well fitted with an equation that takes into account the group delay dispersion difference of HHG between the reference and target paths and the relative carrier-envelope phase change due to the Gouy phase near the focus (see Appendices C & D for the detailed descriptions) [32,33]. This result indicates that the reference and target samples have the same high harmonic emission phase with respect to the MIR driving field.

### B. Floquet state dynamics in thin-film $WSe_2$

Next, with the results for GaSe in hand, we focused on HH interferometry in bulk $WSe_2$. In $WSe_2$, excitons with binding energies of several hundred meV have a dominant contribution to the optical response near the band edge [34]. The previous study using high harmonic spectroscopy suggests that the diabatic and adiabatic transitions between Floquet states of the exciton in $WSe_2$ are induced by intense MIR pulse driving and their signature is imprinted in the high harmonic spectrum [28]. However, the previous result was based on a time-averaged measurement and presented only indirect evidence of the Floquet state dynamics. To clarify Floquet state formation during the pulse duration, it is crucial to access its real-time evolution. In the previous study, a gradual spectral shift of the seventh harmonic into a coherent exciton emission in the time domain was predicted to be direct evidence of the Floquet state dynamics [28].

To confirm the above prediction, we used a mechanically exfoliated 100-nm-thick bulk 2H-WSe$_2$ sample as the target sample and set the MIR intensity at the focus to be 50 GW/cm$^2$ in vacuum. Figure 4A shows the HH intensity spectra of the target bulk WSe$_2$ and the reference GaSe. Since bulk 2H-WSe$_2$ is a centrosymmetric material, the generation of even harmonics are forbidden. Compared with the seventh harmonic spectrum in GaSe, that of WSe$_2$ has a spectral component on the lower energy side of the seventh harmonic energy (1.77 eV) toward the excitonic resonance energy $\varepsilon_X$ = 1.65 eV at room temperature [35]. This spectral feature is attributed to the Floquet state dynamics of the excitons [28]. Figure 4B shows the HH interferogram as a function of photon energy and optical delay. We observed a clear interference signal of odd-order harmonics up to the ninth harmonic. By fitting the data using Eq. (2) and calibrating for the group delay dispersion effect using the results for the GaSe-GaSe combination, we obtained the HH amplitude and phase spectra of WSe$_2$, as shown in Fig. 5A (gray area and blue solid lines, respectively). Here, the emission timing of the fifth harmonics is set to be the time origin for both GaSe-GaSe and WSe$_2$-GaSe combinations so that the derivative of the fifth harmonic phase with respect to the emission energy ($d\phi/d\omega$) becomes zero.

There are two features in HH-phase spectra in WSe$_2$ that differ from those in GaSe (orange dashed lines). One is that the phase spectra of the seventh and ninth harmonics are tilted with respect to the emission photon energy in contrast to the fifth harmonics. These results indicate that the seventh and ninth harmonic emissions are delayed with respect to the fifth harmonic emission, suggesting that the electron dynamics are not instantaneous with respect to the mid-infrared driving field. Another striking feature is that $d\phi/d\omega$ varies within the seventh harmonic spectrum with increasing photon energy. This feature represents that the seventh harmonic emission is chirped, i.e., the lower energy side of the seventh harmonic emission is delayed with respect to the higher energy side.

The simple two-level model considering the excitonic resonance can reproduce the main characteristics of the observed seventh harmonics. Here, we set the transition energy between the two levels as 1.65 eV, the Rabi energy as 0.45 eV, dephasing rate as 0.025 eV, and pulse duration (full width of half maximum) as 53 femtoseconds. The dynamics of the two-level system under these simulation conditions are well understood in terms of the instantaneous Floquet state basis [28,36]. Details of the numerical calculation are described in Ref. 28. Figure 5B shows the simulated HH-amplitude (gray area) and -phase (dark blue dotted lines) spectra. The simulation reproduces the behavior of the fifth and seventh harmonics. Note that the averaged value of the ninth harmonic phase of our experiment is different from the simulated result. In the ninth harmonic emission, where the emission energy is above the bandgap energy, the HHG mechanism is thought to be dominated by recollision of driven electron-hole pairs [37]. A more realistic model that takes into account the details of the band structure of $WSe_2$ may explain the obtained results.

To access the real-time dynamics of the Floquet states in more detail, we reconstructed the temporal profile of the high harmonics based on the experimentally obtained amplitude and phase of the fifth, seventh, and ninth harmonics (see Appendix E). Figures 5C and 5D show the Gabor transforms of the experimental and simulated HH responses, respectively. The experimental results capture the features of the two-level simulation: the delayed response of the seventh harmonic with respect to the fifth harmonic, and a gradual conversion of the seventh harmonic into a coherent exciton emission. This good agreement between experiment and simulation justifies that the following exciton dynamics occur in $WSe_2$ under an intense MIR field (see also Fig. 1A in Ref. 28). First, the initial vacuum state (no excitons) adiabatically transforms into the "dressed" vacuum state and its quasi-energy is gradually shifted with increasing MIR field. Second, at maximum MIR field, the "dressed" vacuum state diabatically transforms into the "dressed" exciton described by Floquet-Landau-Zener tunneling [36]. This diabatic transition creates a superposition state of "dressed" vacuum and "dressed" exciton ("dressed" exciton coherence). Finally, the created dressed exciton coherence emits radiation with its quasi-energy, leading to a gradual energy shift in the emission energy from the seventh harmonic to the bare excitonic resonance as the MIR field decreases.

Our HH-interferometry technique revealed that the high harmonic response near the band edge in $WSe_2$ can be well described by a simple two-level model taking into account the dressed states of the vacuum and exciton.

## V. CONCLUSION

In summary, we have proposed and demonstrated the HH interferometer for real-time observation of high harmonics in solids. We were able to observe clear interference between HH emissions from two different samples in the near-infrared and visible region. We achieved a relative time resolution of less than 150 attoseconds, which is sufficient to follow the sub-cycle electron dynamics. Using the developed HH interferometer, we reconstructed time-resolved HH emissions in bulk WSe$_2$ and revealed the unique time-energy structure resulting from the Floquet state dynamics of the excitons.

Our study provides a convenient method with which to access sub-cycle laser dynamics on a sub-femtosecond scale by using only an infrared laser field for excitation. Note that there are two drawbacks reminded in our measurement system; one is that we can only determine the relative timing of the HH emission, and the other is that spectral bandwidth we can measure is limited by that of reference HH. The former can be complemented with the broadband SFG method of the previous studies [24,25]. Precise real-time characterization of high-photon-flux HHG using SFG provides us with an ideal reference sample for HH interferometry. The combination of these two methods allows us to capture the real-time charge dynamics in quantum materials, which show relatively weaker HH signals. The latter problem can be solved by using a single pulse for driving [38], which has sufficient spectral bandwidth. We believe that our measurement scheme will lead to direct observations of electron trajectories in k-space reflecting band structures and non-trivial topology [39,40,17], scattering effects on charge dynamics [41-43], or strong correlation effects in extreme nonlinear optics [14,16,18,21, 44].

**ACKNOWLEDGEMENTS**

K. U. are thankful to Dr. K. Nagai and Dr. S. Tani for fruitful discussions about HH interferometry. K. U. are also thankful to Dr. T. N. Ikeda and Dr. N. Yoshikawa for their suggestion about the importance to capture the real-time dynamics of the Floquet state. This work was supported by Grants-in-Aid for Scientific Research (S) (Grant No. JP21H05017) and Grants-in-Aid for Scientific Research (C) (Grant No. JP 22K03484). K. U. is thankful for a Grant-in-Aid for Challenging Research (Pioneering) (Grant No. 22K18322).

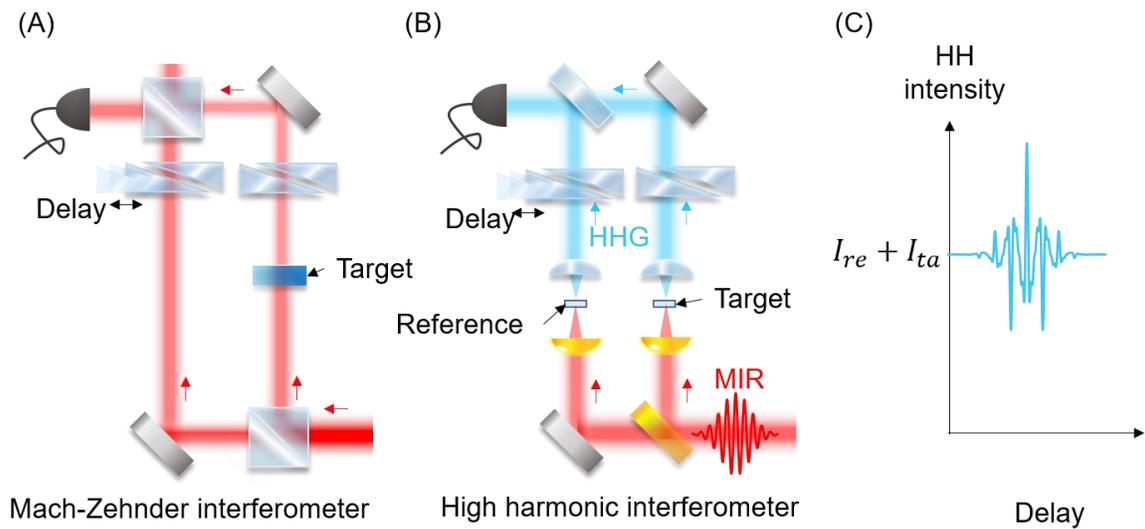

FIG. 1. Principle of HH interferometry. (A) Schematic diagram of Mach-Zehnder interferometer. Here, the optical delay is tuned by inserting a wedge plate. (B) Schematic diagram of Mach-Zehnder interferometer modified for HH interferometry. Here, a reference sample is needed to obtain the interference signal of the HH emissions. (C) Expected HH interferogram obtained by HH interferometer depicted in (B).

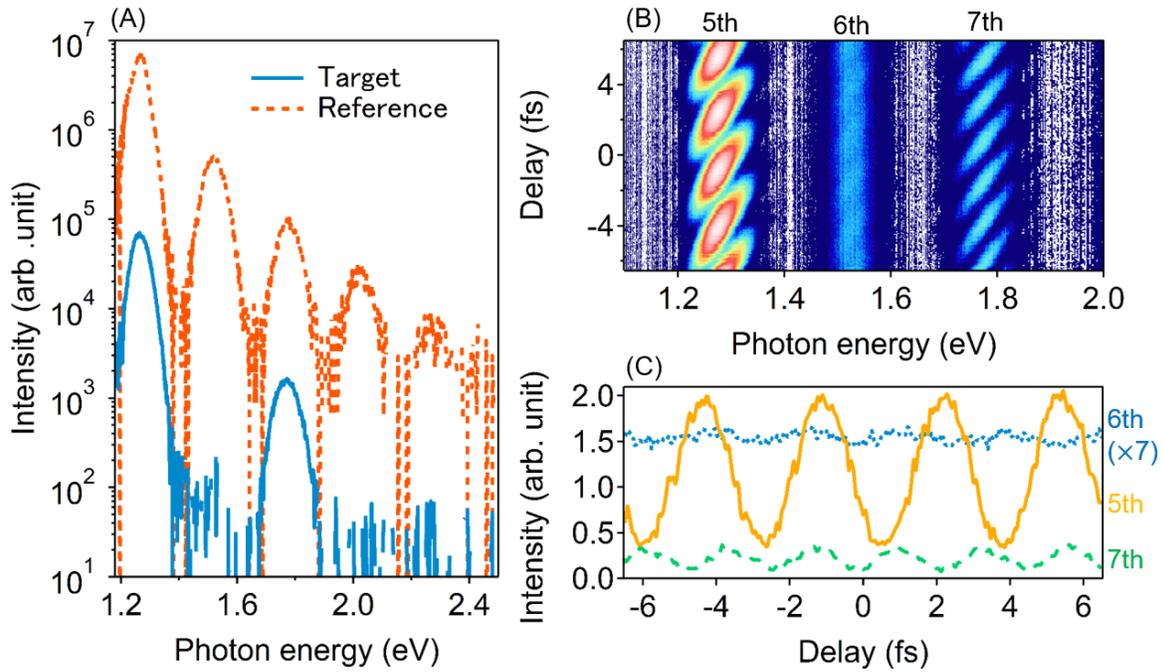

FIG. 2. HH interferometry with GaSe-GaSe reference-target combination. (A) HH intensity spectra of t100-μm-thick GaSe reference (orange dashed line) and mechanically exfoliated GaSe target sample on fused silica substrate (blue solid line). For clarity, the HH intensity of the reference sample is magnified 100 times. (B) Measured HH interferogram as a function of the optical delay and emission photon energy. (C) HH intensities sliced at the center energy of each harmonic (fifth: yellow solid line, sixth: blue dotted line, seventh: green dashed line). Here, the intensity of the sixth harmonic is averaged over the range of its center energies ± 10 meV and magnified 7 times.

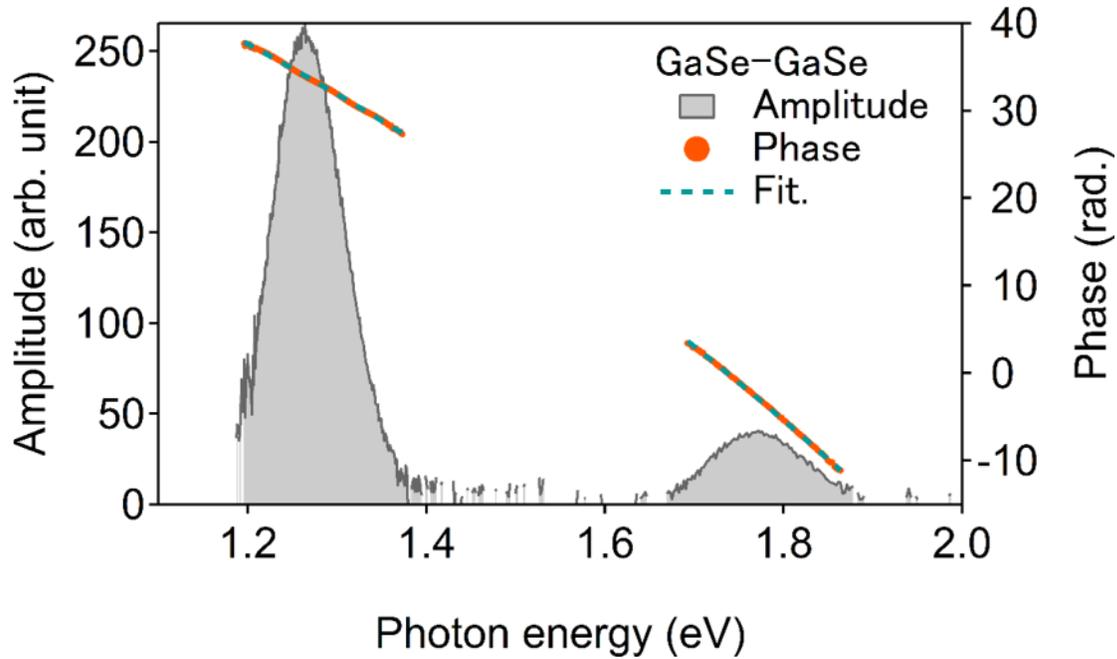

FIG. 3. HH-amplitude (gray shaded area) and -phase (orange solid lines) spectra for a GaSe-GaSe reference-target combination. The blue-green dashed line indicates the fitting result assuming an instantaneous HH response of GaSe, group delay dispersion accumulated in the fused silica wedge plate and substrate for the target sample, and the Gouy phase shift at the focus. In this fitting, the difference in thickness of the fused silica between the reference and target path is estimated to be 0.7 mm. The relative CEP phase difference due to the Gouy phase shift is estimated to be 1.2 radian. See the detailed fitting procedure in Appendix D.

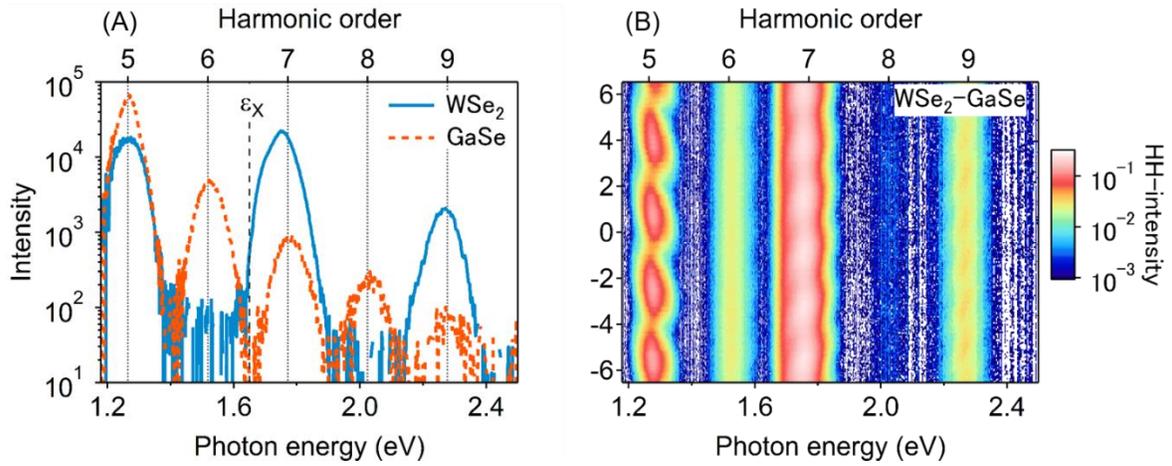

FIG. 4. HH interferometry for a GaSe-WSe$_2$ reference-target combination. (A) HH intensity spectra of the 100-μm-thick GaSe reference (orange dashed line) and mechanically exfoliated WSe$_2$ target on fused silica substrate (blue solid line). The resonance energy of the 1s A-exciton is labeled as $\varepsilon_{exc}$. (B) Measured HH interferogram as a function of the optical delay and emission photon energy.

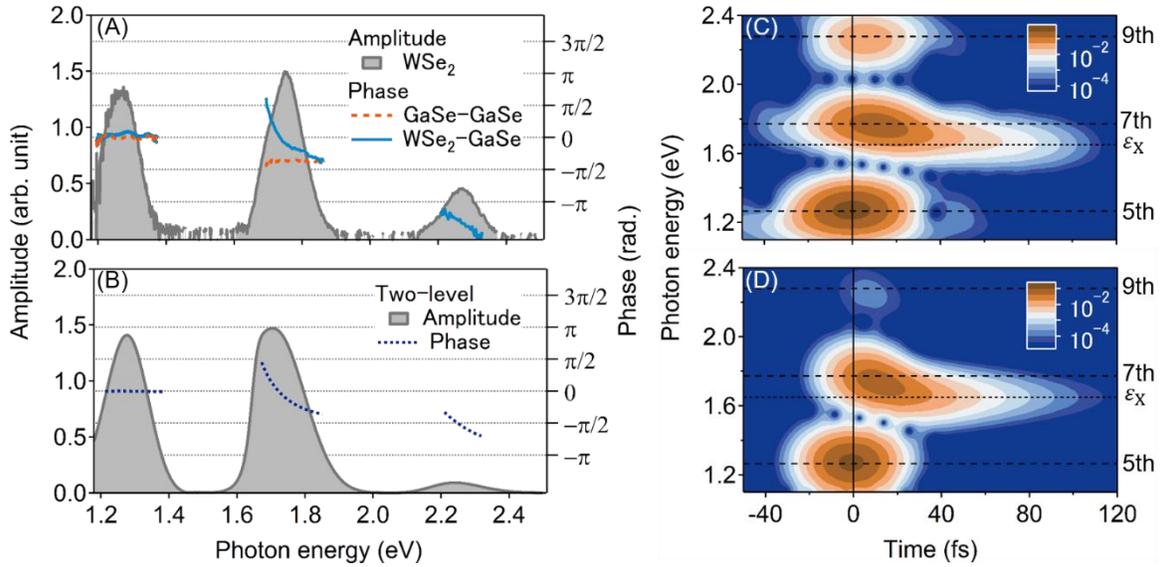

FIG. 5. Floquet state dynamics in WSe$_2$ captured by HH interferometry. (A) Experimentally obtained HH amplitude (gray shaded area) and phase (blue solid lines) spectra of WSe$_2$ as a function of photon energy. For reference, the phase spectrum obtained from the GaSe-GaSe reference-target combination is plotted (orange dashed lines). Here, the effect of group delay dispersion is subtracted from the HH phases by using the fitting result in Fig. 3. The time origin is set to coincide with the fifth harmonic emission peak in the time domain. (B) Numerically simulated HH amplitude (gray shaded area) and phase (dark blue dotted lines) of two-level system considering excitonic resonance corresponding to (A). (C) HH intensity as a function of the emission energy and time obtained by a Gabor transformation of the HH time profile reconstructed from the experimental result (see Appendix. E). (D) Simulated HH intensity as a function of emission energy and time obtained by the Gabor transformation. Black solid, dotted, and dashed lines in (C) and (D) indicate the time origin, excitonic resonance energy ($\varepsilon_X$), and harmonic energies, respectively.

## Appendix A: Detailed experimental setup

The MIR pulses for the excitation were generated by difference frequency mixing of two signal outputs from a dual optical parametric amplifier so that the carrier envelope phase (CEP) of the MIR pulses would be locked. The MIR photon energy $\hbar\Omega$ was 0.253 eV (4900 nm), and the pulse width (full width of half maximum) was estimated to be 100 femtosecond by using the electro-optic sampling. A pair of wire-grid polarizers was used to tune the intensity of the incident MIR pulse.

Figure A1 shows the experimental setup of HH interferometry. MIR pulses are split into two paths by a ZnSe beam splitter, and the separated beams are focused on the reference and target samples by ZnSe lenses. The MIR spot size at the focus was estimated to be 28 μm × 53 μm (FWHM) by using a knife-edge measurement. The high harmonic emissions from each sample are collimated by a $CaF_2$ lens and then combined together by using a sapphire plate as a beam combiner. Here, we chose a combination of the transmitted target signal and the reflected reference signal from a sapphire plate for detection because the transmittance of sapphire is much larger than its reflectance in the visible range, which allows us to suppress the loss of the target signal caused by the beam splitter. To tune the optical delay, we vary the thickness of a pair of fused silica wedge plates in the reference path. To compensate for the effect of the group delay dispersion on HH interferometry as much as possible, we insert the same transmitting optics in the two paths. For example, to compensate for the target signal phase accumulated by passing through the sapphire beam splitter, we insert the same plate in the reference path. The residual group delay dispersion is corrected by the result of HH interferometry in GaSe (see Appendix. D).

## Appendix B: Stability of the HH interferometer.

Figure B1 shows the interferogram of the fifth harmonics for the GaSe-WSe$_2$ reference-target combination as a function of measurement time and delay without any stabilization technique. During the measurement, the time origin of the interferogram experiences a drift, as depicted by the dark blue circles in Fig. B1. The maximum change in the time origin within a 6-hour measurement is estimated to be around 600 attoseconds. Since the fluctuation of the time origin is much very small, it is expected that the time resolution would improve to be less than 100 attoseconds during a long measurement (~ 10 hrs.) by introducing several stabilization techniques such as slow feedback to the interferometer.

## Appendix C: Effect of Gouy phase shift on HH-interferometry.

Focusing a Gaussian beam causes a geometrical phase shift of the wavefront, called the Gouy phase, with respect to the position along the propagation direction of the beam as depicted in Fig. C1 (A). The Gouy phase is given by

$$\phi_{Gouy} = \tan^{-1}\left(\frac{z}{z_R}\right), \tag{C1}$$

where $z_R$ is the Rayleigh length of the focused beam.

When high harmonics are generated with a multi-cycle laser field, the Gouy phase shift of the MIR field causes a phase shift in the HH emission given by

$$\phi_{n,\text{Gouy}} = n \tan^{-1}\left(\frac{z}{z_R}\right), \tag{C2}$$

where $n$ represents the harmonic order [32]. This equation indicates that the HH-phase spectrum shows a step-like behavior depending on the sample position with respect to the focus, as depicted in Fig. C1(B).

To evaluate the effect of the Gouy phase shift on the HH interferometer, we measured the HH-interferogram as a function of the target position along the MIR beam propagation without changing the reference position. Figure C1 (C) shows the phase spectra of the fifth and seventh harmonic emissions for a GaSe-GaSe combination after correcting for the group delay dispersion effect (see Appendix D). By changing the position of the target GaSe sample, the phase offset shifts depending on the harmonic order. Figure C2 (D) shows the HH phase at the center energies of the fifth (solid circles) and seventh (open squares) harmonics as a function of the target sample position. The obtained data are well fitted with Eq. (C2) with $z_R = 445\ \mu m$, as shown by the red solid and yellow dotted curves. From the fitting result, the rate of the phase shift near the focus is estimated to be 10 mrad/$\mu m$ for the fifth harmonic, and 16 mrad/$\mu m$ for the seventh harmonic. For example, to suppress the effect of the Gouy phase shift to less than 100 attoseconds in HH interferometry, we need to locate both the reference and target samples with a precision and accuracy as fine as 15 $\mu m$.

# Appendix D: Correction of the residual group delay dispersion effect in HH-interferometer.

Here, we consider the effect of the group delay dispersion of optics used in the HH interferometer. The electric fields of the HH emissions from the target and reference samples at the position just after the sapphire beam combiner are written as

$$E_{tgt}(t) = \int \frac{d\omega}{2\pi} |\tilde{E}_{tgt}(\omega)| e^{i\left(\omega t + \phi_{tgt}(\omega) + \phi_{n,Gouy}^{tgt} + \phi_{path}^{tgt}(\omega)\right)} . \quad (D1)$$

$$E_{ref}(t) = \int \frac{d\omega}{2\pi} |\tilde{E}_{ref}(\omega)| e^{i\left(\omega t + \phi_{ref}(\omega) + \phi_{n,Gouy}^{ref} + \phi_{path}^{ref}(\omega)\right)}. \quad (D2)$$

Here, label $i$ represents the target($i = tgt$) or reference ($i = ref$) paths, $\tilde{E}_i(\omega)$ is the Fourier transform of $E_i(t)$, $\phi_i(\omega)$ is the HH phase with respect to the driving MIR field, $\phi_{n,Gouy}^i$ is the effect of the Gouy phase shift of the $n$th-order harmonics depending on the sample position in each path given by Eq. (C2). $\phi_{path}^i(\omega)$ represents the phase accumulated by the optical path length and is given by

$$\phi_{path}^i(\omega) = -\int_0^{L_i} \frac{\omega}{c} n_i(\omega, z_i) dz_i, \quad (D3)$$

where the coordinate $z_i$ is defined along the beam propagation, $L_i$ is the path length, $n_i(\omega, z_i)$ is the refractive index as a function of position. $n_i(\omega, z_i)$ contains the information of optics (refractive index, thickness) in each path.

The HH interferogram measured by a spectrometer equipped with a CCD camera is given by

$$I_{\text{HH}}(\omega) = \frac{1}{2}\varepsilon_0 c \left(\tilde{E}_{ref}^*(\omega)\tilde{E}_{tgt}(\omega) + \tilde{E}_{tgt}^*(\omega)\tilde{E}_{ref}(\omega)\right)$$
$$= \varepsilon_0 c |\tilde{E}_{tar}\tilde{E}_{ref}| \cos(\Delta\phi(\omega) + \Delta\phi_{n,Gouy} + \Delta\phi_{path}(\omega)), \quad (D4)$$

where $\Delta\phi(\omega) = \phi_{tar}(\omega) - \phi_{ref}(\omega)$, $\Delta\phi_{n,Gouy} = \phi_{n,Gouy}^{tgt} - \phi_{n,Gouy}^{ref}$, and $\Delta\phi_{path}(\omega) = \phi_{path}^{tgt}(\omega) - \phi_{path}^{ref}(\omega)$.

To perform HH-interferometry, we need to change the optical delay between the target and reference HH emissions. Here, we change the optical delay by tuning the insertion depth of a pair of fused silica wedge plates in the reference path, i.e., $\Delta\phi_{path}(\omega)$. Since we used the same optics for HH beam in both the reference and target paths, $\Delta\phi_{path}(\omega)$ is simplified by using the path length difference $\Delta L$ and the thickness difference of the wedge plates $\Delta L_{wedge}$ as follows:

$$\Delta\phi_{path}(\omega) = -\int_0^{L_{tgt}} \frac{\omega}{c} n_{tgt}(\omega, z_{tgt}) dz_{tgt} + \int_0^{L_{ref}} \frac{\omega}{c} n_{ref}(\omega, z_{ref}) dz_{ref}$$
$$= -\frac{\omega}{c}\Delta L - \frac{\omega}{c}(n_{fused}(\omega) - 1)\Delta L_{wedge}$$
$$= -k(\omega)\Delta L - \Delta k_{fused}(\omega)\Delta L_{wedge}, \quad (D5)$$

where $n_{fused}(\omega)$ is the refractive index of fused silica and $c$ is the speed of light. In our setup, there is a small path length difference between the reference and the target ($|\Delta L| < 1$ mm). To compensate for the phase due to path length difference, we set the finite offset to $\Delta L_{wedge}$ so that $-\Delta k_{fused}(\omega)\Delta L_{wedge}$ is nearly equal to $k(\omega)\Delta L$, where $\Delta L_{wedge} = \Delta L_0 - \delta l$. Here, $\Delta L_0$ is the offset and $\delta l$ is the change relative to the offset to tune the optical delay. Then, $\Delta\phi_{path}(\omega)$ can be rewritten as follows:

$$\Delta\phi_{path}(\omega) = \Delta k_{fused}(\omega)\delta l + \delta\phi_{path}(\omega),$$
$$\delta\phi_{path}(\omega) = -k(\omega)\Delta L - \Delta k_{fused}(\omega)\Delta L_0.$$

The HH-interferogram as a function of $\delta l$ is then given by

$$I_{HH}(\omega, \delta l) \propto \cos\left(\Delta k_{fused}(\omega)\delta l + \Delta\phi(\omega) + \Delta\phi_{n,Gouy} + \delta\phi_{path}(\omega)\right). \quad (D4)$$

This equation corresponds to Eq. (2) in the main text, where the optical delay $\tau$ in Eq. (2) is given by $\tau = \Delta k_{fused}(\omega)\delta l/\omega = (n_{fused}(\omega) - 1)\delta l/c$.

Our purpose in HH-interferometry is to determine the second term $\Delta\phi(\omega)$. The third term $\Delta\phi_{n,Gouy}$ is the relative Gouy phase shift discussed in Appendix. C. The fourth term $\delta\phi_{path}(\omega)$ causes the HH phase spectrum to be modified by the group delay dispersion of the optical wedge plate ($n_{fused}(\omega)$). To evaluate these two additional effects on HH phase spectra, we performed HH interferometry with GaSe-GaSe combinations, as shown in Figs. 2 and 3. By using GaSe samples for both reference and target samples, we assumed that $\phi_{tar}(\omega) \approx \phi_{ref}(\omega)$, and extracted the group delay dispersion in our experimental setup.

We performed a fit of the experimentally obtained HH phase spectrum in Fig. 3 using the following equation:

$$\phi(\omega) = \Delta\phi_{n,Gouy} + \delta\phi_{path}(\omega). \tag{D5}$$

Using the refractive index of fused silica [33], we could obtain a good fit to the experimental data as shown in Fig. 3, yielding $\Delta L = -0.3$ mm, $\Delta L_0 = 0.7$ mm, and $\Delta\phi_{7,Gouy} - \Delta\phi_{5,Gouy} = 1.2\ rad.$.

This good fit indicates that our assumption of $\phi_{tar}(\omega) \approx \phi_{ref}(\omega)$ is valid in our experimental condition, and allowed us to remove the effect of group delay dispersion in the fused silica wedge plate $(n_{fused}(\omega) - 1)\Delta L_0/c$ from the HH phase with GaSe-GaSe and GaSe-WSe$_2$ combinations in Fig. 5(A) and Fig. C1(A).

## **Appendix E: Reconstruction of the HH temporal profiles.**

Here, we describe how to reconstruct the temporal profile of high harmonic emissions. The temporal profile of target HH emission is described by using the HH intensity and phase spectra as follows.

$$E_{\text{tgt}}(t) = \sum_i \delta\omega \left|\tilde{E}_{tgt}(\omega_i)\right| \cos(\omega_i t + \phi_{tgt}(\omega_i)). \tag{E1}$$

For the amplitude of the HH emission $\left|\tilde{E}_{tgt}(\omega_i)\right|$, we used the square root of the HH-intensity spectrum obtained without a reference HH beam:

$$\left|\tilde{E}_{tgt}(\omega_i)\right| = \sqrt{2I_{tgt}(\omega_i)/c\varepsilon_0} \tag{E2}$$

For the phase spectrum of the HH emission $\phi_{tgt}(\omega_i)$, we use the relative phase spectrum $\Delta\phi(\omega_i)$ obtained from HH interferometry under the assumption of $\phi_{ref}(\omega_i) = 0$ as follows:

$$\phi_{tgt}(\omega_i) \approx \Delta\phi(\omega_i) \tag{E3}$$

This assumption is justified when the HH response of the reference sample is instantaneous and the CEP phase of the MIR field is zero. The instantaneous response of the HH emission from a thick GaSe sample was measured in the previous study [24]. If the CEP phase of the driving MIR electric field is given by $\phi_0$, an additional phase of $n\phi_0$ should be added to the $n$th order harmonic phase in Eq. (E3). Note that HH interferometry can only access the relative CEP phase between the target and reference, which originates from the Gouy phase shift, as discussed in Appendix. C. This leads to an uncertainty in the CEP phase of the HH emission. In our measurement, where the driving MIR pulse is a multi-cycle one, the CEP-dependent HH response is negligible. For HH interferometry using a few-cycle MIR pulse, where CEP-dependent HHG is expected, it may be necessary to measure the absolute CEP phase at the sample position by using electro-optic sampling.

Figure E1 (A) shows the reconstructed HH intensity as a function of time in the emission energy range between 1.2 eV and 2.5 eV (including the fifth, seventh, and ninth harmonic emission). The HH intensity shows a temporal periodicity corresponding to that of the MIR field, which is a characteristic of high harmonic generation. Figure E1(B) shows the HH field decomposed into its harmonic components (the fifth, seventh, and ninth harmonics) for clarity. The peaks of the seventh and ninth harmonics are delayed by about 10 femtoseconds compared with that of the fifth harmonic. In addition, the seventh harmonic oscillation lasts much longer than the fifth and the ninth harmonics. This is the signature of the resonance effect of the exciton on the seventh harmonic emission [28].

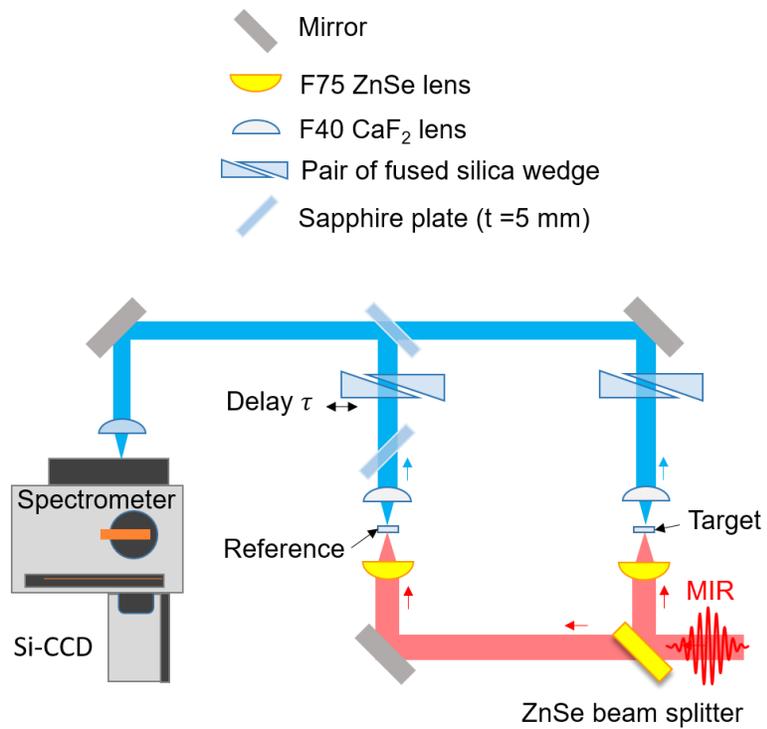

FIG. A1. Experimental setup of the HH interferometer.

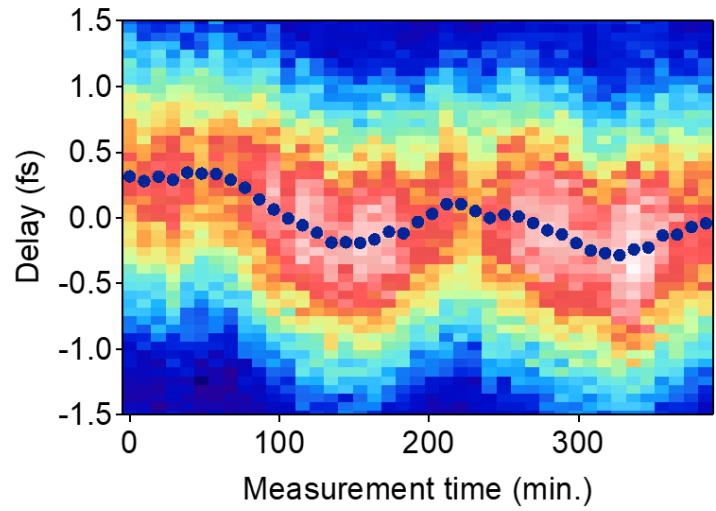

FIG. B1. Long-time stability of HH interferometry. Here, we plotted repeated measurement results of the interferogram at the fifth harmonic energy $I_{HH}(5\Omega, \tau)$ as a function of measurement time. Dark blue circles indicate the delay corresponding to the peak of the interferogram.

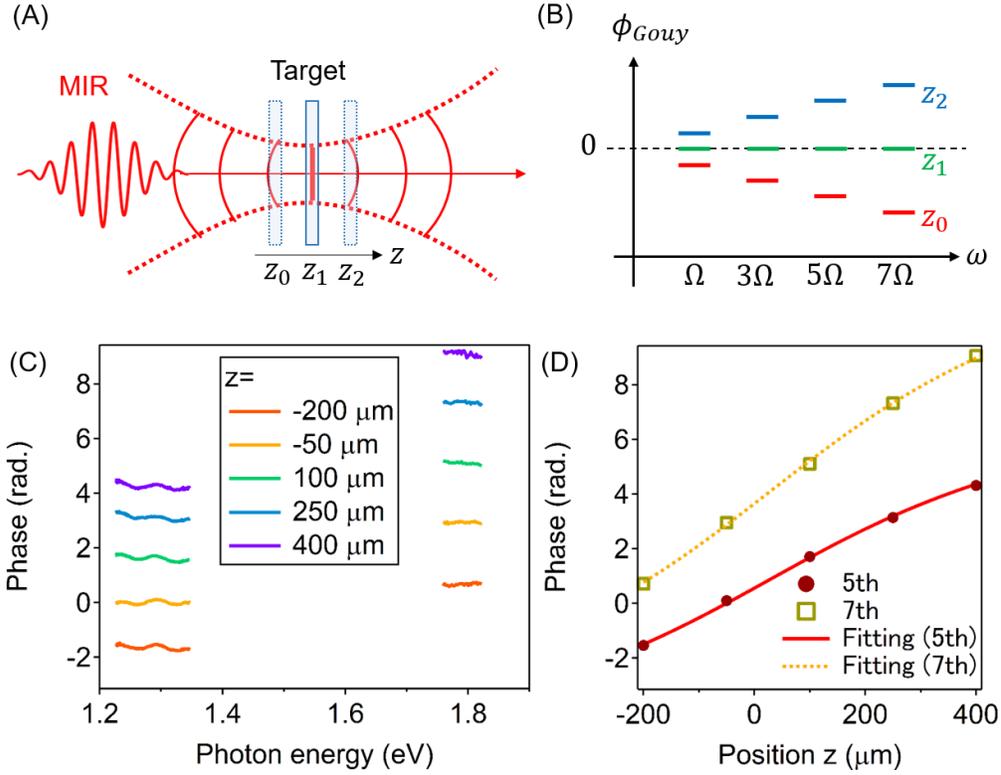

FIG. C1. Effect of the Gouy phase shift in HH interferometry. (A) Schematic diagram of the experimental setup used to measure the Gouy phase shift. The wavefronts of the Gaussian beam (red solid line) are shifted from those of the plane wave. (B) The expected phase spectrum of the high harmonics as a function of the target sample position. (C) Measured HH-phase spectra for GaSe-GaSe reference-target combination for different target positions z. (D) HH phase as a function of the target position z. Dark red circles and yellow open squares indicate the phases of the fifth and seventh harmonics, respectively. Solid red and yellow dotted lines are fitting results using Eq. (C2).

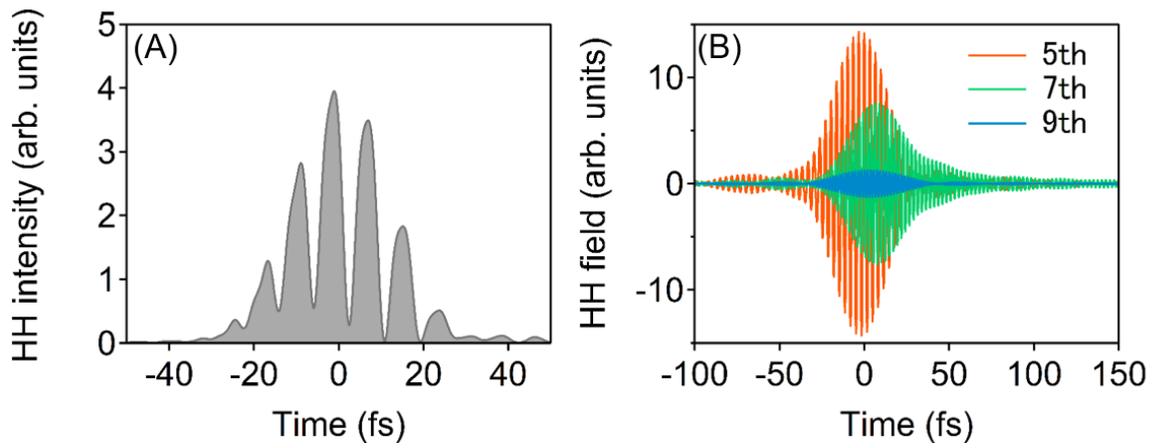

FIG. E1. (A) Temporal profile of the HH intensity in the emission energy range from 1.2 to 2.5 eV (including the fifth, seventh, and ninth harmonic emission). (B) Temporal profiles of the HH field that are decomposed into individual harmonic components (orange: fifth, green: seventh, blue: ninth).